\documentclass[twocolumn,aps,showkeys,showpacs]{revtex4}
\usepackage{amsmath}
\usepackage{graphicx}
\usepackage{epsfig}
\usepackage{color}

\begin{document}

\title{On low-sampling-rate Kramers-Moyal coefficients}
\author{C. Anteneodo}
\affiliation{Department of Physics, PUC-Rio and \\
National Institute of Science and Technology for Complex Systems \\
Rua Marqu\^es de S\~ao Vicente 225, G\'avea, CEP 22453-900 RJ, Rio de
Janeiro, Brazil }
\author{S.M. Duarte Queir\'{o}s}
\affiliation{Centro de F\'{\i}sica do Porto \\
Rua do Campo Alegre, 687, 4169-007 Porto, Portugal}

\begin{abstract}
We analyze the impact of the sampling interval on the estimation of
Kramers-Moyal coefficients. We obtain the finite-time expressions of these
coefficients for several standard processes. We also analyze extreme
situations such as the independence and no-fluctuation limits that
constitute useful references. Our results aim at aiding the proper
extraction of information in data-driven analysis.
\end{abstract}

\date{\today}
\maketitle

\section{Introduction}

The study of systems composed of a large number of degrees of freedom was
endowed a definitive means of reasoning with the introduction of the
differential stochastic dynamical framework about a century ago
\cite{hanggi100}. Specifically, differential stochastic dynamics allows one to
describe the time evolution of a given observable as a function of
macroscopic variables of the system as well as non-deterministic ones. The
latter are expressed in terms of noise(s), reflecting the microscopic
features of the system, whose univocal description is beyond the bounds of
possibility~\cite{risken,gardiner}. Despite the impossibility of a
deterministic description of the evolution of the observable, one can
successfully obtain time dependent statistical details. Undeniably,
Einstein's and Bachelier's grounbreaking works, respectively on Brownian
motion \cite{einstein} and stock price movements \cite{bachelier}, are
outstanding examples of the relevance of differential stochastic dynamics in
diverse fields.

The inference of a differential stochastic process is generically made from
statistical features such as statistical moments, correlation functions and
probability density functions built from a time series of measurements. In
particular, a well established theoretical background for identifying the
stochastic dynamics exists for the important class of processes following
the Markov property~\cite{risken,gardiner}. However, in practice, several
hindrances arise. Specifically, besides matters related to
the finite size of data sets~\cite{mourik} and to
the direct error of the measurements (associated with the quality of the
equipments and/or their calibration~\cite{noise}),
the sample rating (the spell between logged measurements)
plays a crucial role in the determination of actual underlying stochastic
process. As a matter of fact, we are generally left a set of snapshots
reproducing a fraction of the events that occurred in the continuous time
process from which the dynamics is due to be determined. In any case, it
would be important to compare the acquisition interval with the
characteristic times of the process to determine whether the Kramers-Moyal
(KM) coefficients estimates are trustworthy. In this respect, another
difficulty is that the characteristic times are not always known beforehand,
although they can be estimated, e.g., through the computation of
auto-correlation functions. Moreover, when the sampling interval is found to
be too long compared with the characteristic timescales, for suitably
uncovering the process, it is not always possible to upgrade it,
particularly for historical data. For such cases, it is therefore of primary
importance a careful analysis of the impact of the time interval $\tau$ of
data sampling on the observable finite-time KM coefficients, specially as
one approaches the independence limit.

For a Markovian timeseries, one can obtain the evolution equation for the
conditioned probability density (PDF) by computing the KM coefficients,
\begin{equation}
D_{k}(x_{0})=\lim_{\tau \rightarrow 0}\tilde{D}_{k}(x_{0},\tau )\,,
\label{D}
\end{equation}
with
\begin{eqnarray}
\tilde{D}_{k}(x_{0},\tau ) &=&\frac{1}{k!\tau }\int dx P(x,\tau
|x_{0},0)[x(\tau )-x_{0}]^{k}\,  \notag  \label{Dtilde} \\
&\equiv &\frac{1}{k!\tau }\langle \lbrack x(\tau )-x_{0}]^{k}\rangle \,,
\notag \\
&=&\frac{1}{k!\tau }\sum_{j=0}^{k}\left(
\begin{array}{c}
k\cr j
\end{array}
\right) \langle x^{j}\rangle (-x_{0})^{k-j}\, ,
\end{eqnarray}
where we have already assumed that the coefficients are (at least locally)
stationary (for non-stationary data-sets see Ref.~\cite{mourik}).
Aiming to simplify the notation, we denote\ the statistical
averages conditioned to the initial value $x_{0}$: $\langle \cdots \rangle
|_{x=x_{0}}\equiv \langle \cdots \rangle $, while we will reserve $\langle
\cdots \rangle _{u}$ for usual stationary unconditioned averages.

In practice, only the finite-time estimates $\tilde{D}_{k}(x_{0},\tau )$ can
be directly computed, with $\tau $ limited by the minimal time interval,
$\tau _{min}$, of data acquisition. Furthermore, when the sampling interval
$\tau $ is larger than the correlation time, i.e., the PDF $P(x,\tau |x_{0},0)
$ becomes unconditioned and under stationarity, one gets,
\begin{eqnarray}  \label{Dk_independence}
\tilde{D}_{k}^{indep}(x_{0},\tau ) &=&\frac{1}{k!\tau }\int dx
P(x)[x-x_{0}]^{k}  \notag \\
&\equiv &\frac{1}{k!\tau }\langle \lbrack x(\tau )-x_{0}]^{k}\rangle_u \,,
\notag \\
&=&\frac{1}{k!\tau }\sum_{j=0}^{k}\left(
\begin{array}{c}
k\cr j
\end{array}
\right) \langle x^{j}\rangle _{u}(-x_{0})^{k-j}\,.
\end{eqnarray}
Notice that, in Eq.~(\ref{Dk_independence}) the stationary averages are unconditioned,
hence do not depend on $x_{0}$, differently from Eq.~(\ref{Dtilde}).
Therefore, Eq.~(\ref{Dk_independence}) represents a $k$-order polynomial in
$x_{0}$, no matter how complex the intrinsic coefficients $D_{k}$ are. In
practical applications, this feature introduces a complete uncertainty on
the actual form of the intrinsic coefficients, in the absence of a model
\textit{a priori}~\cite{jstat10}.

In this manuscript we obtain the finite-time estimates for several standard
processes described by It\^{o}-Langevin stochastic differential equations,
\begin{equation}
dx=D_{1}(x)dt\,+\,\sqrt{2D_{2}(x)}\,dW_{t}\,,
\end{equation}
where $W_{t}$ is a standard Wiener process. In previous work~\cite{exact},
this task was accomplished for the particular class of processes with
$D_{1}=-ax$ and $D_{2}=Ax^{2}+C$, by means of It\^{o}-Taylor expansions.
Herein, we will consider a larger class of processes following two different
approaches. The first one, based on the solution of the evolution equation
of the raw statistical moments, represents a simplification regarding the
whole process of appraising the finite-time coefficients when the drift is
linear. Second, we employ the Fokker-Planck adjoint operator technique that
is useful for processes with nonlinear drifts, for instance.

Furthermore, we will analyze the extreme instances of no-fluctuations and
independence. The deterministic limit (absence of fluctuations) is relevant
in stochastic dynamical systems since it provides a reference on the
functional behavior of quantities that can be perturbed by augmenting the
intensity (broadness) of the noise. In the meanwhile, the independence limit
sets the bounds expected for large acquisition interval and/or large noise
intensity.

Although it is not always possible a direct inversion of the finite-time
estimates to extract the intrinsic ones, the expressions herein obtained can
be contrasted against the observed KM coefficients, allowing the
identification of the underlying process as well as the associated parameter
values.

\section{Estimating conditional moments}

Given the Fokker-Planck equation (FPE) for $P\equiv P(x,t|x_{0},0)$
\begin{equation}
\partial _{t}P=-\partial _{x}[D_{1}P]+\partial _{xx}[D_{2}P]\,,  \label{FPE}
\end{equation}
the evolution equation for a mean value can be obtained by multiplying both
sides of Eq.~(\ref{FPE}) times the quantity to be averaged and integrating
by parts with suitable boundary conditions (vanishing at the boundaries). In
this way, the equations for the moments have the form
\begin{equation}
\frac{d\langle x^{n}\rangle }{dt}=n\langle x^{n-1}D_{1}(x)\rangle
+n(n-1)\langle x^{n-2}D_{2}(x)\rangle \,.  \label{eqMn}
\end{equation}
If $D_{1}(x)$ is linear and $D_{2}(x)$ is at most quadratic, the equations
for the moments can be successively solved from the lowest order
$n=1$~\cite{CMgenerator}. However, a hierarchy of equations depending on higher order
moments generally arises. In such cases, one can still resort to approximate
techniques such as hierarchy truncation, substitution of terms, etc.
Alternatively, the conditional averages can be computed by means of the
short-time solution expansion~\cite{risken}, which leads to
\begin{equation}
\langle Q(x)\rangle (x_{0},\tau )=\sum_{n\geq 0}[L^{\dag
}(x)]^{n}Q(x)_{|x_{0}}\,\tau ^{n}/n!\,,  \label{adjoint}
\end{equation}
where $L^{\dag }(x)=D_{1}\partial _{x}+D_{2}\partial _{xx}$ is the
(backwards) Fokker-Planck adjoint operator (see also ~\cite{comments}).

In the deterministic case, it is enough to solve Eq.~(\ref{eqMn}) for the
first moment, which becomes
\begin{equation}
\frac{d\langle x\rangle^{det}}{dt}=D_{1}(\langle x\rangle^{det} )\,,
\label{determ}
\end{equation}
where the superindex stands for ``deterministic'' and whence we obtain the
full range of values, $\langle x^{k}\rangle ^{det}=(\langle x\rangle^{det})^{k}$.
Then, the deterministic part of $\tilde{D}_{k}$, with $k\geq 1$, is
\begin{equation}
\tilde{D}_{k}^{det}=\frac{1}{k!\tau }(\langle x\rangle ^{det}-x_{0})^{k}=
\frac{\tau ^{k-1}}{k!}(\tilde{D}_{1}^{det})^{k}\,.  \label{Dk_determ}
\end{equation}
Let us note that the deterministic $\tilde{D}_{k}$ are non-null functions,
although they vanish in the limit $\tau \rightarrow 0$ except for $k=1$.
Nevertheless, we should bear in mind that for not too small $\tau$ the
deterministic part can dominate over the noise contribution. This represents
another drawback one may face in a practical application. In fact, if the
noise is small, the deterministic contribution to $\tilde{D}_{k}$ may screen
the information on the noise.

\section{Linear drift}

Let us first consider the case of linear drift,
\begin{equation}
D_{1}(x)=-ax+b\,,  \label{D1}
\end{equation}
with $a>0$ assuring the existence of a stationary solution. As a matter of
fact, a raft of phenomena are described by an exponential relaxation, which
is concomitant with a parabolic drift potential \cite{gardiner}.
From Eq.~(\ref{eqMn}) one has
\begin{equation}
\frac{d\langle x\rangle }{dt}=-a\langle x\rangle +b\,,  \label{eqM1}
\end{equation}
whose solution is
\begin{equation}
\langle x(\tau )\rangle =x_{0}\,z\,+\,\frac{b}{a}(1-z)\,,  \label{M1}
\end{equation}
assuming the initial condition $\langle x(0)\rangle =x_{0}$ and defining
$z\equiv \mathrm{e}^{-a\,\tau }$. Hence, according to Eq.~(\ref{Dtilde}),
\begin{equation}
\tilde{D}_{1}(x_{0},\tau )=-\biggr(x_{0}-\frac{b}{a}\biggr)\frac{1-z}{\tau }
\,,  \label{D1linear}
\end{equation}
that in the limit $\tau \rightarrow 0$ recovers $D_{1}$. The opposite limit
of independence yields $-(x_{0}-b/a)/\tau $ in accordance with Eq.~(\ref{Dk_independence}).

It is noteworthy that Eq.~(\ref{D1linear}) is independent of the particular
form of $D_{2}$. This is due to the fact that in (and only in) the linear
case, Eq.~(\ref{eqM1}) coincides with its deterministic version,
Eq.~(\ref{determ}), hence $\tilde{D}_{1}=\tilde{D}_{1}^{det}$. For other forms of the
deterministic part of the stochastic differential equation, the derivative
of $\left\langle x\right\rangle $ with respect to the time is equal to more
complex expressions which depend on higher-order moments and/or powers of
the first conditional moment. Despite the slope evolution with $\tau$, it is
worth noting that the finite-$\tau$ drift preserves the linear form of the
intrinsic $D_{1}$.

Equation~(\ref{M1}) also allows one to estimate the conditional two-time
covariance function given by~\cite{gardiner},
\begin{eqnarray}
K(\tau ,x_{0}) &=&\int dx\,x\,x_{0}\,P(x,t|x_{0},0)=x_{0}\langle x(\tau)\rangle  \notag \\
&=&x_{0}^{2}\,z\,+\,x_{0}\frac{b}{a}(1-z)\,,  \label{K0}
\end{eqnarray}
that exponentially decays towards the long-time average. Still one has to
average over initial conditions $x_{0}$ to find,
\begin{equation*}
K(\tau )=\int dxdx_{0}\,xx_{0}P(x,t;x_{0},0)=\int dx_{0}K(\tau
,x_{0})P(x_{0}),
\end{equation*}
where $P(x_{0})$ is usually the steady PDF. However, any choice of $P(x_0)$
will not alter the exponential character of the correlations since the
integration is not carried out over time. In addition, the correlations are
only ruled by the drift and do not depend on the noise amplitude. It is
straightforward to understand that this quirk comes to pass because of the
functional properties of the conditional average.

\subsubsection{Second and higher order conditional moments for quadratic
noise intensity}

If together with drift linearity the noise intensity is second-order
polynomial,
\begin{equation}
D_{2}(x)=Ax^{2}+Bx+C\,,  \label{D2}
\end{equation}
with parameters such that $D_2(x)>0$, then each evolution Eq.~(\ref{eqMn})
is linearly coupled to those of lower order only. In this case the equations
can be solved one after another. In particular, for the second moment, one
has,
\begin{equation}
\frac{d\langle x^{2}\rangle }{dt}=2\left( -(a-A)\langle x^{2}\rangle
+(b+B)\langle x\rangle +C\right) \,.  \label{eqM2}
\end{equation}
Thus, we just need to solve a non-homogeneous 1st order differential
equation of the form
\begin{equation}
f^{\prime }=\hat{A}f+\hat{B}\mathrm{e}^{-\alpha t}+\hat{C}\,,  \label{ode}
\end{equation}
whose solution, given the initial condition $f(0)$, is
\begin{eqnarray}
f(t) &=&\frac{\alpha \hat{C}+\hat{A}(\hat{B}+\hat{C}+(\alpha +\hat{A})f(0))%
} {\hat{A}(\alpha +\hat{A})}\;\mathrm{e}^{\hat{A}t}  \notag  \label{solution}
\\
&&-\frac{\hat{B}}{\alpha +\hat{A}}\mathrm{e}^{-\alpha t}-\frac{\hat{C}}{\hat{%
A}}\,.
\end{eqnarray}

Therefore, $\langle x^{2}(\tau )\rangle =f(\tau )$ with the identifications,
\begin{equation*}
\hat{A}=-2(a-A),
\end{equation*}
\begin{equation*}
\hat{B}=2(b+B)(x_{0}-b/a),
\end{equation*}
\begin{equation*}
\hat{C}=2(b+B)b/a+2C,
\end{equation*}
and $\alpha =a$, following Eqs.~(\ref{eqM2}) and (\ref{M1}), and the initial
condition $f(0)=x_{0}^{2}$. Namely,
\begin{eqnarray}
&&\langle x^{2}(\tau )\rangle =z^{2}wx_{0}^{2}+2\frac{b+B}{a-2A}
(z-z^{2}w)x_{0}  \notag  \label{M2} \\
&&+\,\frac{b(b+B)}{a(a-A)(a-2A)}(az^{2}w-2(a-A)z+a-2A)  \notag \\
&&+\frac{C}{a-A}(1-z^{2}w)\,\,,
\end{eqnarray}
where $z\equiv \mathrm{e}^{-a\tau }$ and $w\equiv \mathrm{e}^{2A\tau }$.

The finite-$\tau$ second KM coefficient can be obtained by means of
Eq.~(\ref{Dtilde}) with $\langle x(\tau)\rangle$ and $\langle x^2(\tau)\rangle$ given
by Eqs.~(\ref{M1}) and (\ref{M2}), respectively. Explicitly, we find
\begin{eqnarray}  \label{D2linear}
&&\tilde{D}_{2}(x_{0},\tau ) =\frac{1}{2\,\tau } \biggl( [ 1-2z+z^2 w ]
\,x_{0}^{2}  \notag \\
&& + 2\left[ \frac{b+B}{a-2A} ( z-z^2w) +\frac{b}{a} ( z-1) \right] \,x_{0}
\notag \\
&& +\frac{b(b+B)(a z^2w -2(a-A) z +a-2A)}{a(a-A)(a-2A)}  \notag \\
&&+\frac{C}{a-A}(1-z^2w) \biggr) \,.
\end{eqnarray}

When $\tau \rightarrow 0$, $\tilde{D} _{2}\rightarrow D_{2}$, while for
$a\tau >>1$, Eq.~(\ref{Dk_independence}) is verified.

These results permit one to embrace many fundamental models such as
Ornstein-Uhlenbeck, Feller and harmonic drift with
additive-(linear)multiplicative noise. Also in finance, in most well-known
models of volatility ($\sigma =\sqrt{x}$), the drift is linear and
$D_{2}=D|x|^{\beta }$, with $\beta =$ 0 (Ornstein-Uhlenbeck process), 1
(square-root, Feller or Cox-Ingersoll--Ross model) and 2 (Hull and White
model).

Let us also recall that for the linear drift, its finite-time expressions
are also linear in $x_{0}$. Similarly, with linear drift and $\beta=0,1,2$
the dependence of $\tilde{D}_{2}$ on $x_{0}$ is quadratic. As a matter of
fact, a similar scenario holds for $kth$-order finite-time KM coefficients.
In other words, the $kth$-order KM coefficients are polynomials of order $k$,
whose coefficients depend on $\tau$. Thus, it is not simple to nimbly
rescue the value of $\beta$ just by identifying the polynomial order of the
observed coefficients.

For the quadratic noise intensity, the evolution equations of higher order
conditional moments, from Eq.~(\ref{eqMn}), are
\begin{equation}  \label{ODE}
\frac{d\left\langle x^{n}\right\rangle }{dt}=A_{n}\left\langle
x^{n}\right\rangle +B_{n}\left\langle x^{n-1}\right\rangle
+C_{n}\left\langle x^{n-2}\right\rangle ,
\end{equation}
where
\begin{eqnarray*}
A_{n} &=&n\left[ \left( n-1\right) A-a\right] ; \\
B_{n} &=&n\left[ \left( n-1\right) B+b\right] ; \\
C_{n} &=&n\left( n-1\right) C.
\end{eqnarray*}

High-order KM coefficients are particularly relevant as well. In other
words, for the present class of Markovian processes the Pawula
theorem~\cite{risken} implies that they must vanish. In the absence of any other reason
to discard the validity of Chapman-Kolmogorov approach, if the finite-time
conditional higher-moments are non-null, it is important to probe whether it
is the outcome of a finite-time effect. Let us restrain our calculations to
the explicit formulae of the third and fourth coefficients in two important
cases: \emph{(i)} the reversion to a null value in the absence of coupling
between additive and (linear) multiplicative white noises ($b=0$ and $B=0$)
and \emph{(ii)} the Feller process.

(i) For the additive-multiplicative process with $b=0$ and $B=0$, notice
that Eq.~(\ref{eqMn}) with $n=3$ has the same form of Eq.~(\ref{ode}),
through the identification $\hat{A}=-3(a-2A)$, $\hat{B}=6Cx_{0}$, $\hat{C}=0$
and $\alpha =a$, and the initial condition $f(0)=x_{0}^{3}$, yielding

\begin{equation}
\langle x^{3}(\tau )\rangle =\frac{3x_{0}C}{a-3A}
(z-(zw)^{3})+x_{0}^{3}(zw)^{3}\,.  \label{M3}
\end{equation}
Then, according to Eq.~(\ref{Dtilde}), one has
\begin{equation}
\tilde{D}_{3}(x_{0},\tau )=\frac{1}{6\tau }\biggl(\langle x^{3}(\tau)
\rangle -3x_{0}\langle x^{2}(\tau )\rangle +3x_{0}^{2}\langle x(\tau)
\rangle -x_{0}^{3}\biggr)\,,  \label{D3linear}
\end{equation}
where the averaged quantities are given by Eqs.~(\ref{M3}), (\ref{M2}) and
(\ref{M1}), respectively.

For the fourth moment, Eq.~(\ref{eqMn}) with $n=4$ has the same form of
Eq.~(\ref{ode}), through the identification $\hat{A}=4(3A-a)$, $\hat{B}
=12C^{2}/(A-a)+12Cx_{0}^{2}$, $\hat{C} =-12C^{2}/(A-a)$ and $\alpha =2(a-A)$,
together with the initial condition $f(0)=x_{0}^{4}$, yielding
\begin{eqnarray}
&&\langle x^{4}(\tau )\rangle =z^{4}w^{6}x_{0}^{4}+\frac{
6C(z^{2}w-z^{4}w^{6})}{a-5A}\,x_{0}^{2}  \notag  \label{M4} \\
&+&3C^{2}\frac{(a-A)z^{4}w^{6}-2(a-3A)z^{2}w+a-5A}{(a-A)(a-3A)(a-5A)}
\,.\;\;\;
\end{eqnarray}
Thus,
\begin{eqnarray}
&&\tilde{D}_{4}(x_{0},\tau )=\frac{1}{24\tau }\biggl(\langle x^{4}(\tau
)\rangle -4x_{0}\langle x^{3}(\tau )\rangle  \notag  \label{D4linear} \\
&&+6x_{0}^{2}\langle x^{2}(\tau )\rangle -4x_{0}^{3}\langle x(\tau )\rangle
+x_{0}^{4}\biggr)\,,
\end{eqnarray}
is obtained by substitution of the averaged quantities Eqs.~(\ref{M4}), (\ref{M3}),
(\ref{M2}) and (\ref{M1}). The present results generalize those
previously obtained for the particular case $D_{1}=-ax$ (with $a>0$) and
$D_{2}=Ax^{2}+C$ in~\cite{exact}.

\begin{figure}[h!]
\centering
\includegraphics*[bb=150 75 510 730, width=0.45\textwidth]{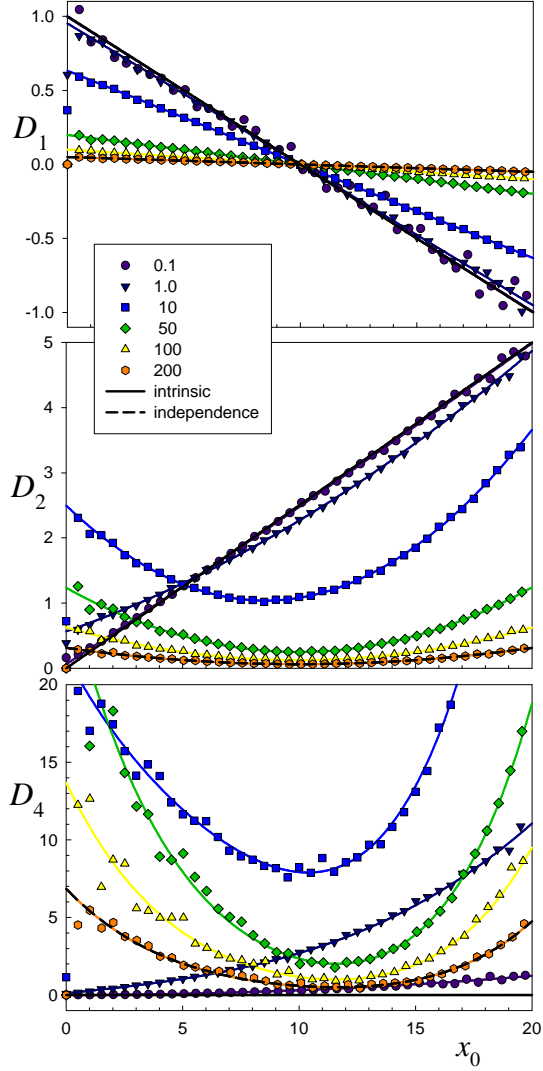}
\caption{Color online. Impact of $\protect\tau$ on first, second and fourth finite-time KM
coefficients for the process $dx=(-ax+b)dt+\protect\sqrt{2Bx}\,dW$, with
$a=0.1$, $b=1.0$ and $B= 0.25$. The values of $\protect\tau$ used are
indicated on the figure. Artificial timeseries were generated by means of
the Euler algorithm with $dt=10^{-3}$ and $10^6$ data points of the
timeseries were considered for the computation of KM coefficients in each
case. Symbols correspond to numerical estimates and the associated
colored solid lines to the theoretical expressions for finite-time coefficients given by
Eqs.~(\protect\ref{D1linear}), (\protect\ref{D2linear}) and
(\protect\ref{D4feller}).
Black solid lines correspond to the intrinsic coefficients $D_1=-ax+b$,
$D_2=Bx$ and $D_4=0$ and black dashed lines (practically coinciding with the
curves for $\tau=200$), to the independence limit forms given by
Eq.~(\protect\ref{Dk_independence}) with $\protect\tau=200$ }
\label{fig:feller}
\end{figure}

(ii) For the Feller process ($A=C=0$), by successive integration of
Eqs.~(\ref{ODE}) and using (\ref{Dtilde}) we find,
\begin{eqnarray}  \label{D3feller}
&& \tilde{D}_3(x_0,\tau) = \frac{(1-z)^3}{6\tau} \left( -x_0^3+3
\frac{b-(b+2B)z}{a(1-z)}x_0^2 \right.  \notag \\
&& \left. - 3 \frac{(b+B)(b-(b+2B)z)}{a^2(1-z)}x_0 + \frac{b(b+B)(b+2B)}{a^3}
\right) \;\;\;\;\;\;\;\;
\end{eqnarray}
and
\begin{eqnarray}  \label{D4feller}
&& \tilde{D}_4(x_0,\tau) = \frac{(1-z)^4}{24\tau} \left( x_0^4 - 4 \frac{%
b-(b+3B)z}{a(1-z)}x_0^3 \right.  \notag \\
&& +6 \frac{ (b+2B)[(b+3B)z-2(b+B)]z + b(b+B) }{a^2(1-z)^2}x_0^2  \notag \\
&& -4\frac{(b+B)(b+2B)(b-(b+3B)z)}{a^3(1-z)}x_0  \notag \\
&& \left.+ \frac{b(b+B)(b+2B)(b+3B)}{a^4} \right).
\end{eqnarray}

\begin{figure}[h!]
\centering
\includegraphics*[bb=150 275 510 730, width=0.45\textwidth]{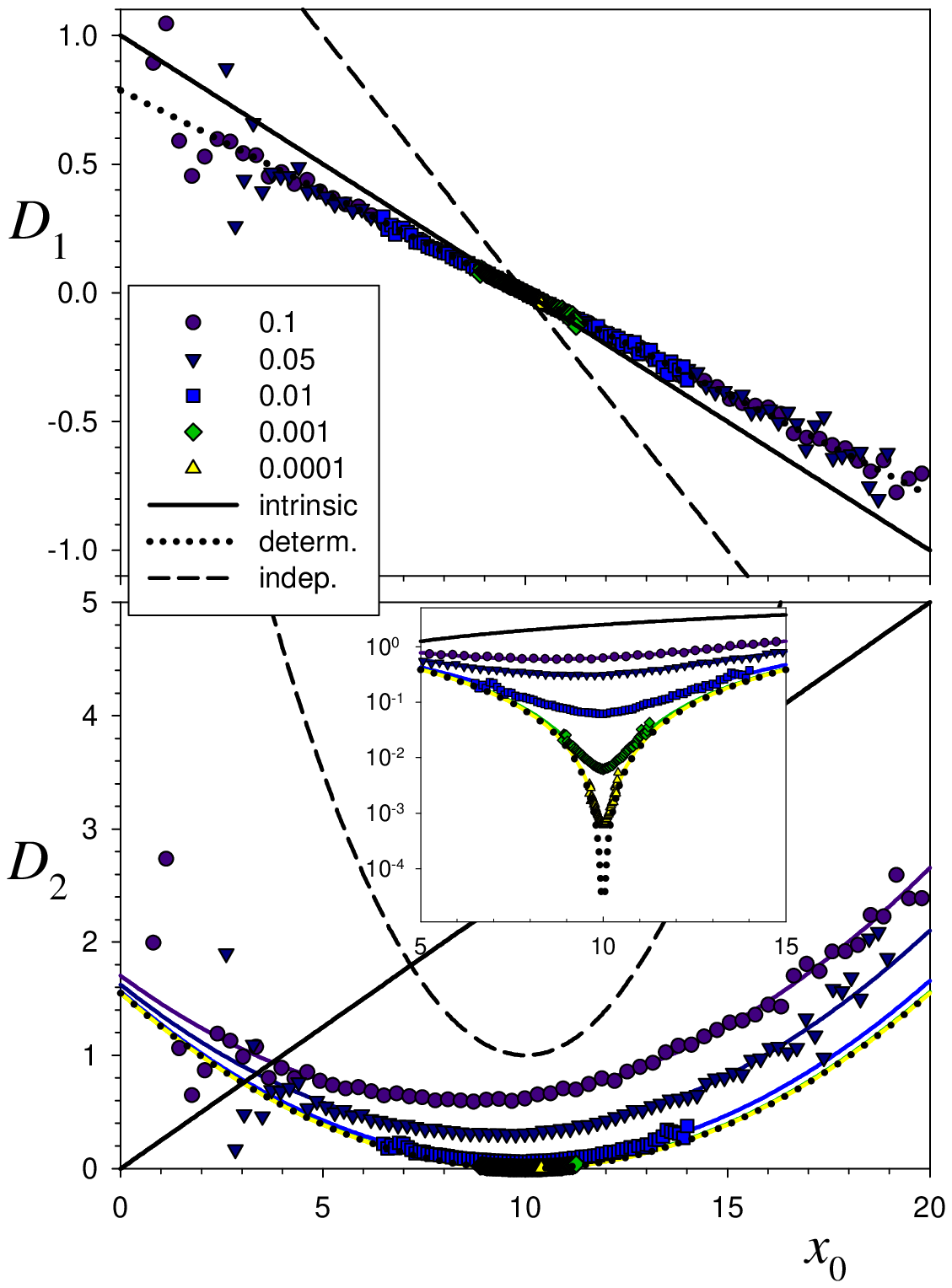 }
\caption{Color online. Impact of the noise intensity on the first and second finite-time
KM coefficients for the same process of Fig.~\protect\ref{fig:feller}. In
all cases $\protect\tau=5$ and the different values of $B$ are indicated on
the figure.
Black solid lines correspond to the intrinsic coefficients,
black dashed lines to the independence limit forms at $\protect\tau=5$ and
black dotted lines to the deterministic forms (Eq.~(\protect\ref{Dk_determ}))
at $\protect\tau=5$.
Symbols correspond to numerical estimates and the associated  colored (thin)
solid lines to the theoretical expressions
for finite-time coefficients given by Eqs.~(\protect\ref{D1linear}) and
(\protect\ref{D2linear}). In the lower panel the inset shows the same data in
log-linear scales. }
\label{fig:fellerDET}
\end{figure}

Considering that for the Feller process $\langle x\rangle^n_u=(B/A)^n
\frac{\Gamma(b/B+n)}{\Gamma(b/B)}$, then the KM coefficients can be rewritten in
terms of the unconditioned raw moments, which can be straightforwardly
estimated from data, as follows,

\begin{equation}
\tilde{D}_1(x_0,\tau) = \frac{(1-z)}{\tau} \langle x -x_0\rangle_u .  \notag
\end{equation}

\begin{equation}
\tilde{D}_2(x_0,\tau) = \frac{(1-z)^2}{2\tau} \left\langle (x-x_0)^2
\right\rangle_u,  \notag
\end{equation}

\begin{eqnarray}
&&\tilde{D}_{3}(x_{0},\tau )=\frac{(1-z)^{3}}{6\tau }\left\langle
\left(x-x_{0}\right) ^{3}\right\rangle _{u}  \notag \\
&&+\frac{z\left( 1-z\right) ^{2}}{\tau } \frac{\sigma _{x}^{2}}{\left\langle
x\right\rangle _{u}^{2}} \left\langle x(x-x_0) \right\rangle _{u} \,x_{0},
\notag
\end{eqnarray}
and
\begin{eqnarray}
&&\tilde{D}_{4}(x_{0},\tau )=\frac{(1-z)^{4}}{24\tau }\left\langle \left(
x-x_{0}\right) ^{4}\right\rangle _{u}  \notag \\
&&+ \frac{z\left( 1-z\right) ^{3}}{2\,\tau }\frac{\sigma _{x}^{2}}{
\left\langle x\right\rangle _{u}^{2}} \left\langle x(x-x_0)^2 \right\rangle
_{u} x_{0}
%\notag \\&&
+\frac{z^{2}\left( 1-z\right) ^{2}}{ 2\,\tau }
\frac{\sigma _{x}^{2}}{\left\langle x\right\rangle _{u}^{2}} x_{0}^{2}  \, \notag
\end{eqnarray}
where $\sigma _{x}^{2}=\left\langle x^{2}\right\rangle _{u}-\left\langle
x\right\rangle _{u}^{2}$.

In Fig.~\ref{fig:feller}, we exemplify the behavior of $\tilde{D}_{1}$,
$\tilde{D}_{2}$ and $\tilde{D}_{4}$ for the Feller process $dx=(-ax+b)\,dt+
\sqrt{2Bx}\,dW$ (that is $A=C=0$, or also $\beta=1$) as the independence
limit is approached. Notice the variety of behaviors that can be observed
even for not too large $\tau$. The slope of the linear $\tilde{D}_1$ changes
with $\tau$ taking the values $-(1-z)/\tau$, while $\tilde{D}_2$ soon
becomes quadratic with $\tau$. Also $\tilde{D}_4$ largely departs from the
intrinsic null value. For all the coefficients, the independence limit is
practically attained at $\tau \simeq 200$ ($\tau >>1/a=10$).

Figure~\ref{fig:fellerDET} illustrates the impact of the intensity of noise
on the finite-time estimates. Noise intensity does not affect $\tilde{D}_1$
which coincides with the deterministic straight line. $\tilde{D}_2$ largely
disagrees with the true form $D_2$, departing towards the independence
limit, for large noise intensity, and towards the deterministic limit for
small noise. Of course, as one approaches the deterministic
(no-fluctuations) limit the range of $x_0$ shrinks around the deterministic
(equilibrium) value.

When the drift is linear, whatever the noise intensity, the parameters $a$
and $b$ can be unraveled by a linear fit to the observed $D_1$: the value
of the slope $-(1-z)/\tau$ allows one to determine the parameter $a$, once
$\tau$ is known, while the abscissa at which $\tilde{D}_1$ vanishes
corresponds to $b/a$. That is, if the observed drift can be considered
linear in good approximation, a direct inversion of $\tilde{D}_1$ to obtain
the intrinsic $D_1$ is possible. This will also facilitate the obtention of
the functional form (and parameter values) of $D_2$.

\subsubsection{Non-quadratic noise intensity}

For $\beta \neq 0,1,2$, or more generally for non-quadratic $D_{2}$, the
finite-time $\tilde{D}_{2}$ are also non quadratic in $x$. The equation for
the second moment, following Eq.~(\ref{eqMn}),
\begin{equation}
\frac{d\langle x^{2}\rangle }{dt}=-2a\langle x^{2}\rangle +2b\langle
x\rangle +2\langle D_{2}(x)\rangle \,  \label{eqM2beta}
\end{equation}
cannot be solved straightforwardly in this case.

Let us consider the especial case with $\beta =3$ (known as 3/2-model) used
as a volatility model as well \cite{embrechts}. For small noise intensity,
one can find the correction $c_{2}$ to the deterministic part of $\langle
x^{2}\rangle $ by means of expansion (\ref{adjoint}), taking $Q(x)=x^{2}$ .
By identifying the general form of the coefficients of $\tau^{n}$, with the
aid of algebraic manipulation programs, we obtain
\begin{eqnarray}
c_{2} &=&\frac{C}{a^{4}}\left(
b^{3}+6b^{2}(ax_{0}-b)z+(3b^{3}-6a^{2}bx_{0}^{2}+2a^{3}x_{0}^{3})z^{2}
\right.  \notag \\
&+&\left. 6ab\tau (ax_{0}-b)^{2}z^{2}-2z^{3}(ax_{0}-b)^{3}\right) \,.
\end{eqnarray}

Then
\begin{equation}
\langle x^{2}\rangle =(\langle x\rangle ^{det})^{2}\,+\,c_{2}\,+\,\mathcal{O}
(C^{2})\,  \label{x2linear}
\end{equation}
and on that account,
\begin{equation}
\tilde{D}_{2}=\frac{\tau }{2}(\tilde{D}_{1}^{det})^{2}\,+\,
\frac{c_{2}}{2\tau }\,+\,\mathcal{O}(C^{2})\,,  \label{D2lowlinear}
\end{equation}
which in the limit $\tau \rightarrow 0$ tends to $D_{2}\equiv Cx^{3}$ as
expected.

\section{Nonlinear drift}

As in the case of linear drift and non-quadratic noise, if the drift is
nonlinear, the equations for the moments cannot be solved successively from
the lowest order. Moreover, the useful independence of the evolution equation
for $\left\langle x\left( t\right) \right\rangle $ on higher order
statistical moments, which are noise dependent, fails. Consequently, the
existence of a nonlinear drift introduces a troublesome relation between
$\tilde{D}_{1}$ and the noise as well as a dependence of the correlation
function on the same noise. However, we can still use the calculations at
the deterministic limit to represent the upper bound which is quite reliable
for the cases presenting small noise intensity. Explicitly, if the
contribution of diffusion is neglected, the evolution is almost
deterministic. Accordingly, we will solve the deterministic equation for the
first moment whereas for higher order moments we will take into account the
lowest order correction due to noise.

Let us consider a process with cubic drift~\cite{siegert} also called
Bernoulli oscillator,
\begin{equation}
dx=(-ax-bx^{3})dt\,+\,\sqrt{2C}\,dW_{t}\,,  \label{SDEcubic}
\end{equation}
where $b,C\geq 0$. When $a<0$ the system represents a stochastic motion in a
bistable potential. The statistical moments of this stochastic system were
early studied within a variational approach context~\cite{phythian}.
However, this method lands up introducing an extra (fitting) parameter which
we adamantly want to avoid, since it would increase the level of uncertainty
of the results.

\begin{figure}[h!]
\centering
\includegraphics*[bb=150 75 510 730, width=0.45\textwidth]{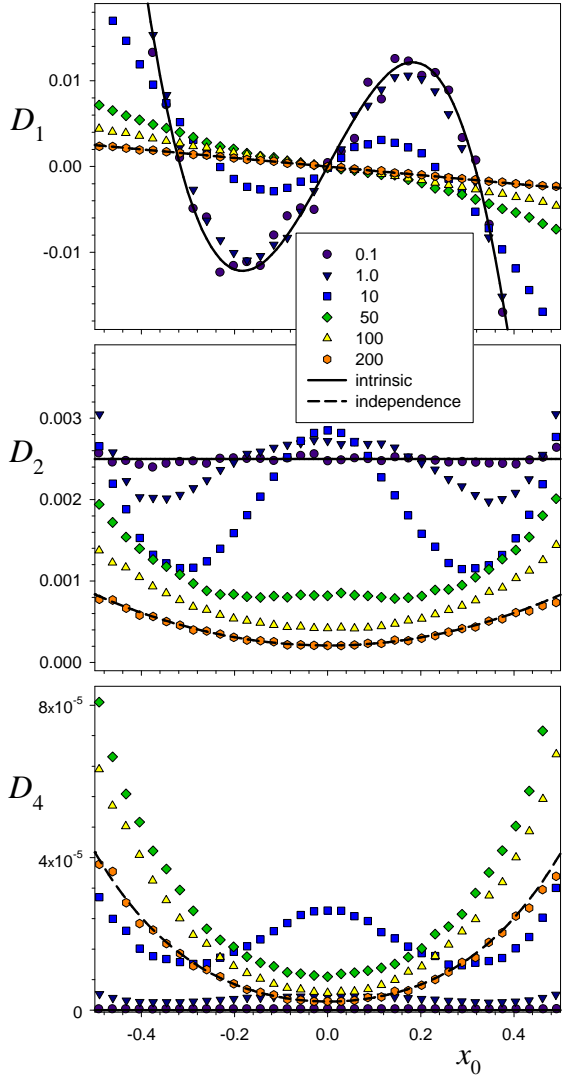}
\caption{Color online. Impact of $\protect\tau$ on the first, second and fourth
finite-time KM coefficients, for the process defined by
Eq.~(\protect\ref{SDEcubic}), with $a=-0.1$, $b=1$ and $C=0.0025$. The values of $\protect\tau
$ are indicated on the figure.
Symbols correspond to numerical estimates. Black solid lines correspond to the intrinsic
coefficients $D_1=-ax-bx^3$, $D_2=C$ and $D_4=0$, while black dashed lines to the
independence limit forms at $\protect\tau=200$. }
\label{fig:cubic}
\end{figure}

Figure~\ref{fig:cubic} exhibits the first, second and fourth finite-time KM
moments, obtained for different values of the sampling interval $\tau$.
Notice that, with increasing $\tau$, $\tilde{D}_1$ tends to a linear form
that can erroneously lead to assume that the drift is linear. However, under
this assumption, the inversion of $\tilde{D}_1$ should give incongruous
results for $D_1$, thus leading to discard the linearity of $D_1$. For small noise
intensity, the deterministic expression  fitted (with two fitting parameters $a$ and
$b$) to the observed $\tilde{D}_1$ allows to recover  $D_1$. Notice
also the complex influence of $D_1$ on the higher order coefficients,
leading to forms with two minima, in contrast with the linear drift case.

\begin{figure}[h!]
\centering
\includegraphics*[bb=150 275 510 730, width=0.45\textwidth]{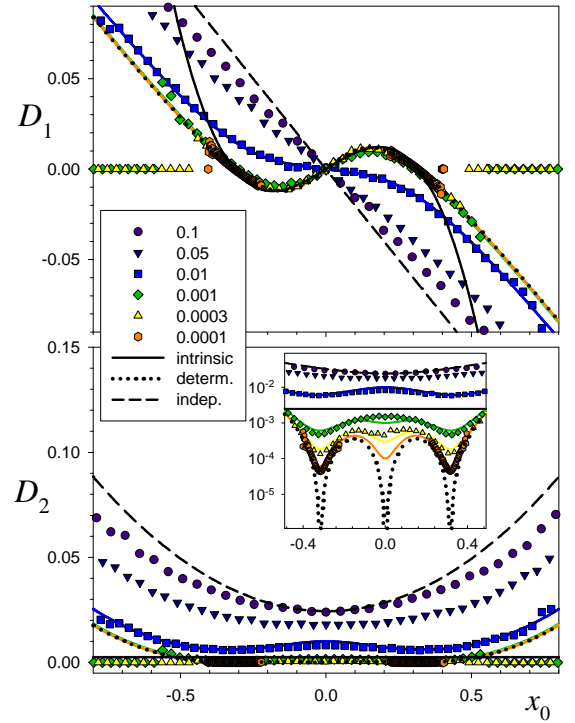}
\caption{Color online. Impact of the noise intensity on the first and second finite-time
KM coefficients, for the same process of Fig.~(\protect\ref{fig:cubic}). In
all cases $\protect\tau=5$, and the different values $C$ used are indicated
on the figure.
Black  solid lines correspond to the intrinsic coefficients,
black dashed lines to the independence limit at $\protect\tau=5$, and black
dotted lines to the deterministic forms (\protect\ref{Dk_determ}) at
$\protect\tau=5$.
Symbols correspond to numerical estimates and the associated colored lines
to the approximations for small
noise intensity given by Eqs.~(\protect\ref{D1low}) and (\protect\ref{D2low}),
respectively. In the lower panel the inset shows the same data in
log-linear scales. For the smallest noise intensity two different initial
conditions were used to fill both potential wells.}
\label{fig:cubicDET}
\end{figure}

By integration of Eq.~(\ref{determ}), one obtains the deterministic
expression,
\begin{equation}
\tilde{D}_{1}^{det}(x_{0},\tau )=\left( \frac{\mathrm{e}^{-a\tau }}
{\sqrt{1+b\,x_{0}^{2}(1-\mathrm{e}^{-2a\tau })/a}}-1\right) \frac{x_{0}}{\tau }\,,
\label{zeroth}
\end{equation}
that does not preserve the original simple cubic form underscoring the
unique character of the linear drift. In the limit $b\rightarrow 0$, one
recovers the linear case studied above. The case where $b$ is a perturbative
parameter may be of interest for systems in the vicinity of a phase
transition, such as in Refs.~\cite{ph_tr}. In the limit $a\rightarrow 0$,
Eq.~(\ref{zeroth}) becomes,
\begin{equation}
\tilde{D}_{1}^{det}(x_{0},\tau )=\left( \frac{1}{\sqrt{1+2b\,x_{0}^{2}\tau }}
-1\right) \frac{x_{0}}{\tau }\,.  \label{purecubic}
\end{equation}
Independent of the signal of $a$, for short $\tau $,
$\tilde{D}_{1}^{det}(x_{0},\tau )$ is equal to $-a\,x-b\,x^{3}$. For large values
of $\tau $, the picture depends on whether $a$ is positive or negative. In the
former case, which corresponds to a single well potential, $\tilde{D}{1}^{det}$
is defined by a straight line,
\begin{equation}
\tilde{D}_{1}^{det}(x_{0},\tau )=-\frac{x_{0}}{\tau },
\end{equation}
which corresponds to the independence limit. Meanwhile, in the latter case,
which entails bi-stable potentials, one obtains,
\begin{equation}
\tilde{D}_{1}^{det}(x_{0},\tau )=\left( \frac{\mathrm{e}^{|a|\,\tau }}{\sqrt{
1+b\,x_{0}^{2}(\mathrm{e}^{2\,|a|\,\tau }-1)/|a|}}-1\right) \frac{x_{0}}{\tau }.
\end{equation}
In this case, for large values of $|a|\,\tau $, one can describe the limits
corresponding to large values of $b/|a|\,x_{0}^{2}e^{2\,|a|\,\tau }$,
\begin{equation}
\tilde{D}_{1}^{det}(x_{0},\tau )\approx \sqrt{\frac{|a|}{b}}\frac{1}{\tau }-
\frac{x_{0}}{\tau },  \label{anegx0big}
\end{equation}
and for small values of the same quantity,
\begin{equation}
\tilde{D}_{1}^{det}(x_{0},\tau )\approx \frac{\exp \left[ |a|\,\tau \right]
-1}{\tau }x_{0}-\frac{b\exp \left[ 3|a|\tau \right] }{2\,|a|\,\tau }x_{0}^{3}.
\end{equation}
As a result, we can verify that the polynomial dependence of
$\tilde{D}_{1}^{det}(x_{0},\tau )$ is preserved for the central region that dwindles
as the sampling rate increases resulting in the straight line limit
(\ref{anegx0big}), which describes the full dependence (beyond relaxation scale)
situation.

In the generic case $D_{1}=-h\,x^{n}$ (with $h,n>0$),
the solution to Eq.~(\ref{determ}) is,
\begin{equation}
\tilde{D}_{1}^{det}(x_{0},\tau )=\left( \frac{1}{(1+(n-1)h\,x_{0}^{n-1}
\tau)^{\frac{1}{n-1}}}-1\right) \frac{x_{0}}{\tau }\,,
\end{equation}
which includes Eq.~(\ref{purecubic}) as a particular case (for $n=3$) and
also the linear case with $b=0$ (for $n=1$). In the latter instance it is
provable that the exponential functional dependence is recovered. For all
the cases and in the limit $\tau \rightarrow 0$, one recovers $D_{1}(x)$
whereas in the opposite limit $\tau \rightarrow \infty$, one gets $-x/\tau $.

For small fluctuations, higher order finite-$\tau $ KM moments are dominated
by the deterministic part given by Eq.~(\ref{Dk_determ}) with further
corrections dependent on the noise.

For the cubic drift, in the presence of small fluctuations, such that $C\sim a$,
\begin{equation}
\langle x\rangle =\langle x\rangle ^{det}+c_{1}+\mathcal{O}
(C^{2},aC,a^{2})\,,  \label{x1cubic}
\end{equation}
with,
\begin{equation}
c_{1}=-Cbx_{0}\tau ^{2}\frac{3+4y+2y^{2}}{(1+2y)^{5/2}}
\end{equation}
and $y=bx_{0}^{2}\tau $. Then,
\begin{equation}
\tilde{D}_{1}=\tilde{D}_{1}^{det}+\frac{c_{1}}{\tau }+\mathcal{O}
(C^{2},aC,a^{2})\,,  \label{D1low}
\end{equation}
where $\tilde{D}_{1}^{det}$ is given by Eq.~(\ref{zeroth}).

Similarly, for the second moment we found the first correction to the
deterministic part
\begin{equation}
\langle x^{2}\rangle =(\langle x\rangle ^{det})^{2}+c_{2}+\mathcal{O}
(C^{2},aC,a^{2})\,,  \label{x2cubic}
\end{equation}
with
\begin{equation}
c_{2}=\frac{2C\tau }{(1+2y)^{3}}\,.
\end{equation}
Then, finally
\begin{equation}
\tilde{D}_{2}=\frac{\tau }{2}(\tilde{D}_{1}^{det})^{2}+
\frac{c_{2}-2x_{0}c_{1}}{2\tau }+\mathcal{O}(C^{2},aC,a^{2})\,,  \label{D2low}
\end{equation}
which verifies that $\tilde{D}_{2}$ tends to $D_{2}\equiv C$ as $\tau
\rightarrow 0$.

Figure~\ref{fig:cubicDET} shows the impact of the intensity of noise on the
finite-time estimates. In contrast to the linear drift case, here the noise
influences $\tilde{D}_{1}$, in such a way that as the intensity of noise $C$
increases, $\tilde{D}_{1}$ approaches the independence limit even for small
sampling rates. This could be attributed to the decorrelating role of noise
(it is worth regarding that $\tilde{D}_{1}$ already depends on the noise for
non-linear potentials). Similarly to the linear case, $\tilde{D}_{2}$
largely disagrees with the true form $D_{2}$, departing towards the
independence limit, for large noise intensity, and towards the deterministic
limit for small noise. Notice that if the intensity of noise is too small,
one of the wells of the potential may become inaccessible. Precisely, let us
focus on the insets in the lower panel of fig.~\ref{fig:cubicDET}. When one
is in the no-fluctuation regime the system is characterized by three fixed
points, namely one unstable (at $x=0$, which prevails if and only if $x=0$
is the initial condition) and two stable points (at $x=\pm \sqrt{a/b}$).
Minimal noise is able to make the observable, which we can envisage as a
virtual particle, move off the unstable point to one of the wells, each one
defined by one of the remaining extremes (minima), but it gives a very
little rate of transition between the regions of stability. As the noise
intensity augments, the rate of transition increases as well and we are able
to detect the finite values of the KM moments in the vicinity of the
unstable point. This can be adduced by computing the first
non-vanishing eigenvalue, $\lambda $,  when the FPE is transformed
into a Schro\"{e}dinger equation \cite{risken}.
In first approximation, the eigenvalue $\lambda $ of
this bistable potential,  is given by,
\begin{equation}
\lambda =\int_{0}^{\infty } dx\,{\rm e}^{ \frac{a}{2\,c}x^{2}+\frac{b}{4\,c}
x^{4}} \int_{x}^{\infty }  dy\,{\rm e}^{
  -\frac{a}{2\,c}y^{2}-\frac{b}{4\,c}y^{4} }.
\end{equation}
For the values presented in Fig. \ref{fig:cubicDET},
$\lambda$ goes from $\lambda \approx 0.2448$ (for $c=0.1$) to
$\lambda \approx 6\times 10^{-13}$ (for $c=10^{-4}$).

\section{Final remarks}

In this manuscript we obtained analytical results about the impact of the
sampling rate on the KM coefficients directly computed from timeseries.

We analyzed stochastic processes subject to a linear drift, which already
comprise a huge variety of processes. We managed to compute exact
expressions for the evolution of conditional moments, from which we obtained
the finite-time KM coefficients and correlation functions. The first moment
and the linear correlation function are both independent of the noise
intensity.

We also analyzed standard nonlinear drift cases, where the scenery exhibits
a sharp change. The moments are now ruled by a cascade of differential
equations which hampers the obtention of exact analytical solutions.
Nonetheless, approximate expressions valid for small noise intensity were
achieved by means of the adjoint operator approach. Differently from the
linear case, the intensity of the noise affects every conditional moment and
the correlation function as well. We also showed that by increasing the
noise intensity the functional dependence of the KM heads towards the
independence regime, which is characterized by a polynomial of the same
order. This fact has notorious consequences in time series analysis, since
one can be brought onto a situation where independent-like coefficients are
obtained, but one is unable to assign this finding to either an improper
sample rating or a strong noise intensity.

For sampling rates smaller than the relaxation time, in a steady state
approach, our results permit one to sieve the set of processes giving raise
to the heuristic stationary probability density function in order to select
the appropriate Markovian differential stochastic process and consequently
to determine the respective parameter values. Furthermore, it enables the
judgement of the validity of the Markovian proposal by supplying precise
estimates of higher order KM moments. Otherwise, if the relaxation time is
shorter than the sampling rate, the measured KM coefficients will tend to a
polynomial of order matching the order of the moment. This functional form
is independent of the primary stochastic process, hence introducing
uncertainty in the recognition of the process.

Uncertainties might be hedged by inspecting probability densities and
correlation functions as well as other measures of dependence such as the
relative (Kullback-Leibler) entropy.

\section*{Acknowledgements:}

CA is grateful to Brazilian agencies Faperj and CNPq for partial financial
support and SMDQ acknowledges the warm hospitality of PUC-Rio and the
financial support of the National Institute of Science and Technology for
Complex Systems during the early stage of this work.

\end{document}